\documentclass[anonymous]{llncs}

\usepackage{booktabs} 
\usepackage{url}
\usepackage{xspace}
\usepackage[font=normal]{caption}
\usepackage{subcaption}

\usepackage{float}
\usepackage{textcomp}
\usepackage{multirow,makecell}
\usepackage{pifont}
\usepackage{tikz}
\usepackage{bbm}
\usepackage{adjustbox}
\newcommand{\etal}{{\em et al.}\xspace}
\newcommand{\ie}{{\em i.e.,}\xspace}
\newcommand{\eg}{{\em e.g.,}\xspace}
\newcommand{\etc}{{etc.}\xspace}
\newcommand{\BfPara}[1]{{\vspace{1.5mm}\noindent\bf#1.}\xspace}
\usepackage{times}

\usepackage{xcolor}

\usepackage{listings}
\usepackage{tikz}
\usetikzlibrary{tikzmark}
\usetikzmarklibrary{listings}
\usepackage{amsmath}
\usepackage{amssymb}

\usepackage{amsmath}
\usepackage{graphicx}
\usepackage{makecell}

\usepackage{colortbl}
\definecolor{darkgreen}{rgb}{0.0, 0.3, 0.13}
\definecolor{darkred}{rgb}{0.2, 0.0, 0.13}
\newcommand{\cc}{\cellcolor{black!5!white}}
\newcommand{\ccd}{\cellcolor{blue!5!white}}

\usepackage{tcolorbox}
\tcbset{textmarker/.style={%
        parbox=false,boxrule=0mm,boxsep=0mm,arc=0mm,
        outer arc=0mm,left=3mm,right=3mm,top=3pt,bottom=3pt,
        toptitle=1mm,bottomtitle=1mm}
        }
        
\newtcolorbox{blueBox}{textmarker,
    colback=blue!10!white}

\usepackage[english]{babel}
\usepackage{blindtext}
\usepackage{xspace}
\usepackage{filecontents}
\usepackage{multicol}

\usepackage{hyperref}

\lstdefinestyle{myStyle}{
  belowcaptionskip=1\baselineskip,
  breaklines=true,
  language=C++,
  showstringspaces=false,
  basicstyle=\footnotesize\ttfamily,
  keywordstyle=\bfseries\color{green!40!black},
  commentstyle=\itshape\color{purple!40!black},
  identifierstyle=\color{blue},
  stringstyle=\color{orange},
  numbers=left,
  firstnumber=1,
}

\newcommand{\stcd}{{\sc StructCoder}\xspace} 
\newcommand{\ours}{{SCAE}\xspace}

\title{Untargeted Code Authorship Evasion with Seq2Seq Transformation}

\author{Soohyeon Choi\inst{1} \and Rhongho Jang\inst{2}  \and DaeHun Nyang\inst{3} \and David Mohaisen\inst{1}}
\institute{University of Central Florida \and Wayne State University  \and Ewha Womans University}
\begin{document}
\maketitle

\begin{abstract}
Code authorship attribution is the problem of identifying authors of programming language codes through the stylistic features in their codes, a topic that recently witnessed significant interest with outstanding performance.
In this work, we present \ours, a code authorship obfuscation technique that leverages a Seq2Seq code transformer called \stcd. \ours customizes \stcd, a system designed initially for function-level code translation from one language to another (\eg Java to C$\#$), using transfer learning. 
\ours{} improved the efficiency at a slight accuracy degradation compared to existing work. We also reduced the processing time by $\approx$ 68\% while maintaining an 85\% transformation success rate and up to 95.77\% evasion success rate in the untargeted setting.


\end{abstract}

\vspace{-7mm}
\section{Introduction}
\vspace{-3mm}
Code authorship attribution identifies the author(s) of a source code written in a particular programming language~\cite{AbuhamadAMN18,IslamHLNVYG15}. 
Several studies have been introduced to address this task, exploiting that code often contains the programmers' stylistic patterns, extracted as distinguishing code features uniquely identifying the code author. Such features may include the style and structure of the code, comments, variable names, and function names and have shown great success in identifying single and multiple authors of the same piece of code~\cite{AbuhamadAMN18,IslamHLNVYG15,AbuhamadANM20}. However, code authorship attribution can be abused. For instance, it would be undesirable that the code authorship attribution technique would not withstand a misleading attempt by attributing the same code to multiple authors~\cite{li2022ropgen}. An adversary capable of false attribution will overcome the forensics effort and defeat malware origin tracking and cyber threat profiling~\cite{MohaisenAM15,AlasmaryAJAANM20,AlasmaryKAPCAAN19}. 

To defeat the attribution techniques, Quiring \etal~\cite{quiring2019misleading} proposed misleading authorship attribution with automatic code transformations (we refer to their approach by MAA). They applied various types of transformation to the source code (\eg changes to control flow, APIs, declarations, \etc) and then used Monte-Carlo Tree Search (MCTS) to choose the optimal transformed code that is syntactically correct, semantically equivalent to the original code, and likely to be misattributed by authorship attribution methods. However, this approach has various limitations. Most noticeably, it has significant memory-compute requirements for finding the optimal solutions among many possibilities. This issue is a fundamental limitation because of the brute-force nature of MCTS in attempting many different options to find the most optimal among them.

Inspired by MAA~\cite{quiring2019misleading} and to overcome some of its limitations, we propose a code authorship obfuscation with an automated sequence-to-sequence (Seq2Seq) model that avoids searching for the best transformation. Our approach utilizes a Seq2Seq-based code authorship evasion technique called \ours{}. We exploited the advantages of Seq2Seq models by customizing \stcd~\cite{tipirneni2022structcoder}, a recently proposed model with state-of-the-art performance in code translation. \stcd is designed and trained for code translation tasks, \ie to translate code written in one programming language to another programming language (\eg Java to C$\#$), and is not meant to transform a source code to a target code in the same language. 
Therefore, we extended \stcd's capabilities to ``transform'' the code in one language to another code in the same language. In doing so, we build a source-target dataset for code transformation tasks formulated as a transfer learning task~\cite{torrey2010transfer}.

\ours{'}s evaluation showed that it reduced the processing time by $\approx$68\% (from $\approx$11,500 minutes to $\approx$3,600 minutes) and achieved 85\% of transformation success and 95.77\% of evasion rates, significantly reducing resources usage and processing time while achieving competitive transformation and evasion performance.

\BfPara{Contributions} Our contributions are summarized as follows: (1) Our source-target pair dataset is built using codes from the code transformation method with MCTS to train a machine learning-based Seq2Seq model. (2) \ours, utilizing a pretrained Seq2Seq model called \stcd, is customized for code transformation, fine-tuned with our source-target pair dataset, and formulated as a transfer learning task for generating transformed codes that are syntactically correct and semantically equivalent to the original code. The produced codes are misclassified by the attribution method at a high rate. (3) \ours{'s} evaluation, showing it reduced the processing time and resource usage compared to the existing work while maintaining (or improving) the performance.

\BfPara{Organization} The organization of the rest of this paper is as follows. The preliminaries are presented in~section~\ref{sec:relatedwork}. Our work's challenges, overview, and technical details are introduced in section~\ref{sec:codeauthorship}.
Our results and analysis are presented in section~\ref{sec:evaluation}. Finally, our conclusions are drawn in section~\ref{sec:conclusion}. 

\begin{figure}[t]
    \centering
    \includegraphics[width=0.9\textwidth]{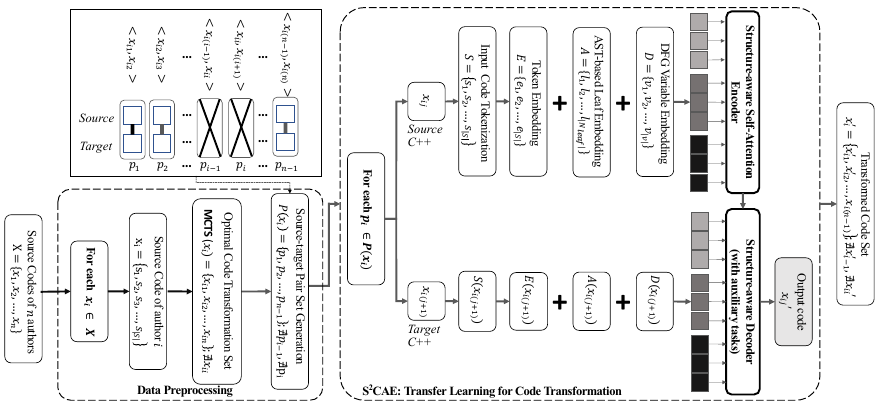}\vspace{-1mm}
    \caption{An illustration of \ours{'s} pipeline.}
    \label{fig:proposedmodel}\vspace{-1mm}
\end{figure}

\section{Our Approach: Building Blocks}\label{sec:relatedwork}
We aim to utilize the Seq2Seq model, \stcd~\cite{tipirneni2022structcoder}, to transform source code into target code in a way that misclassifies the code author. 
This requires training \stcd with source-target code pairs, traditionally obtained through manual transformations, which is challenging (\eg prone to errors and laborious). 
To simplify this process, we employ MCTS as an oracle to generate training data, focusing on producing syntactically and semantically correct code variants with distinct programming styles. 
This allows us to train and fine-tune \stcd for code transformation instead of just translation. 
For evaluation, we rely on Abuhamad \etal's work~\cite{AbuhamadAMN18} as our primary code authorship attribution method due to its strong performance on large-scale datasets previously used for assessing MCTS. 
As a result, we will explain the three methods which we employed in this work in more detail in this section.

\BfPara{Code Authorship Attribution}
The code authorship attribution problem is defined as the task of identifying the programmer~\cite{AbuhamadAMN18} or group of programmers~\cite{AbuhamadANM20} who wrote a piece of code. 
Programming languages (\eg C++, Java) often contain various stylistic patterns of programmers~\cite{AllamanisBBS14}---\eg a programmer may prefer using a {\tt for} loop instead of a {\tt while} loop or generating an output in files instead of printing them on the terminal. These patterns can be used as features to identify authors of source codes, and this task is known as code authorship attribution.

Abuhamad \etal~\cite{AbuhamadAMN18} proposed a system called Deep Learning-based Code Authorship Identification System (DL-CAIS). 
In the data preprocessing phase of their system, features are extracted from source codes using a Term Frequency - Inverse Document Frequency (TF-IDF) vectorizer. These features are then used to train a deep learning model with three recurrent neural networks (RNNs) and three fully connected layers (FCLs) to capture distinctive code features. To address the challenge of identifying authors in cases with numerous authors, deep representations are employed to build a robust random forest (RF) model for author classification.
As a result, DL-CAIS achieved an accuracy of 92.3\% with over 8.9K authors and was used to identify authors of four different programming languages, C, C++, Java, and Python. 

\BfPara{Misleading Authorship Attribution}
Authorship attribution poses a privacy risk by potentially revealing an author's identity, making it a concern for those seeking anonymity. Conversely, it can also be exploited to falsely assign credit or blame for code creation. For instance, attackers might use such techniques to shift focus away from their involvement in code development by attributing a bug or vulnerability to someone else. These diverse possibilities have driven research into misleading authorship attribution.

Quiring \etal~\cite{quiring2019misleading} proposed misleading authorship attribution with automatic code transformations. MAA performed a series of semantics-preserving code transformations to deceive DL-CAIS. Quiring \etal~\cite{quiring2019misleading} considered the code transformation as a game against the code authorship attribution and utilized MCTS to guide the transformation in the feature space to find a short sequence of transformations.
With MCTS, the transformer can find a transformed code that is syntactically correct and semantically equivalent to the original code by minimal modification. 
However, MCTS can be computationally intensive, as it requires simulating many possible moves and evaluating their outcomes. Thus, the authors attempted to decrease computation time by limiting the number of transformations, implementing an early stop technique, \etc Despite their efforts, this method still requires more resources compared to others.

\BfPara{Seq2Seq Model: \stcd}
Machine learning is widely used to analyze, understand, and generate natural language~\cite{khan2016survey,aone1998trainable,le2021machine,oliinyk2020propaganda,ayanouz2020smart}.  However, utilizing machine learning methods for programming languages presents difficulties due to factors such as syntax and organization, terminology, and the surrounding context.
Programming languages have strict rules for syntax and structure (\eg ``;'' of C families and indentations of Python language), and even small errors can cause a program to fail. 
The vocabulary of a programming language is typically much smaller than the vocabulary of a natural language. This means that there is less data available for training machine learning models, which can be a challenge. 
The context in which a piece of code is used can have a significant impact on its meaning. This can make it more difficult to analyze code and understand its purpose.
Despite these challenges, researchers have been developing machine learning techniques that can be applied to programs for various tasks~\cite{ahmad2021unified,tipirneni2022structcoder,phan2021cotext}. 

Most recently, Tipirneni \etal~\cite{tipirneni2022structcoder} proposed a transformer-based encoder-decoder model called \stcd, a structure-aware transformer for text-to-code generation and code translation. They used a structure-aware self-attention framework for their model and pretrained it using a structure-based denoising autoencoding task. 
As a result, the model is trained to generate a source code based on a given text or translate a source code into a different language (\eg Java $\leftrightarrow$ C$\#$). 
In addition, \stcd uses the T5 transformer~\cite{raffel2020exploring}, augmented for modeling code structure, source syntax, and data flow when given a source code and generates a code that is syntactically correct and semantically equivalent to the source code. 
To apply this structure-aware transformer for code translation, the authors trained their model with 10K samples, tested with 1K samples, and achieved an 88.41 CodeBLEU score in Java $\rightarrow$ C$\#$ translations, which is the highest score on a code translation task with CodeXGLUE benchmark~\cite{CodeXGLUE}.

\section{\ours{}: Putting it Together}\label{sec:codeauthorship}
We transfer the MCTS task to Seq2Seq learning for fast and efficient code authorship evasion. The proposed \ours{} extracts the functional behavior of the MCTS model and then transfers the knowledge to a lightweight Seq2Seq-based programming language processing (PLP) model for code transformation. 
In the following, we first provide a high-level overview of \ours{}'s architecture. Then, we explain the technical details for extending a PLP model's capacity from code translation (different languages) to code transformation (the same language).

\begin{table}[t]
    \centering
    \begin{tabular}{lc}
        \Xhline{2\arrayrulewidth}
                                \ccd{Method}& \ccd{Accuracy}     \\
        \Xhline{2\arrayrulewidth}
         Abuhamad \etal~\cite{AbuhamadAMN18}       &          84.98\%                              \\ 
         \cc{Caliskan-Islam \etal~\cite{IslamHLNVYG15}}        & \cc{90.5\%}                 \\            
            
        \Xhline{2\arrayrulewidth}
\end{tabular}
    \caption{Baseline attribution accuracy.}
    \label{tab:baseline}\vspace{-8mm}
\end{table}

\BfPara{Challenges and Overview}
Figure~\ref{fig:proposedmodel} illustrates the architecture of \ours. 
Given a source code, \ours{} leverages a transfer learning accelerated PLP model, \stcd, to transform (not translate) the code without retaining authors' discriminative features. 
We note that \stcd is designed for translating code to a different language. Thus, we extend the model to code transformation within the same language, leveraging a transfer learning technique and parameters fine-tuned.
As we can see in the transfer learning stage in Figure~\ref{fig:proposedmodel},
the authorship misleading performance of \ours{} is guaranteed by summarizing the functional behavior of MCTS via retrieving source-target code pairs, and then fine-tuning the \stcd model with the code pairs. As such, \ours{} significantly reduces the training and code transformation costs. 

\BfPara{Data Preprocessing} To construct source-target pairs, we used the outputs of MAA~\cite{quiring2019misleading}. Given a set of code samples from $n$ authors, each author's code ($x_i$) is transformed with MCTS and produces $n-1$ transformed codes with different stylistic patterns. We note that $x_{ii}$ is a code that is transformed with the author's own stylistic patterns and, thus, is excluded, \ie $|\text{\sf MCTS}(x_i)|=n-1$. All transformed codes are functionally equivalent to the original code $x_i$, but 
with minimal changes in stylistic patterns to deceive the authorship attribution method. We built a source-target dataset with the transformed code set. The source-target pairs ($\text{\sf P}(x_i)$) of a source code $x_i$ are characterized by the following: $\text{\sf P}(x_i) = \{<x_{i1}, x_{i2}>, <x_{i2}, x_{i3}>, \cdots, <x_{i(n-1)}, x_{in}>\}; \nexists <x_{x(i-1)}, x_{ii}>, \nexists <x_{ii}, x{i(i+1)}>$, where $<x_{ij}, x_{i(j+1)}>$ denotes two adjacent codes in the transformed set as a pair. We note that two pairs that involve $x_{ii}$ are excluded, and thus $|\text{\sf P}(x_i)|=n-2$, where $n$ is the number of stylistic patterns.

\BfPara{Transfer Learning with Pretrained Model} 
We fine-tuned the pre-trained \stcd model with our new dataset for transfer learning~\cite{torrey2010transfer}.
This allows us to transfer \stcd's function from code translation to code transformation, with a relatively small amount of data.  
The success of the technique has been verified in image recognition tasks, \eg using visual geometry group (VGG)~\cite{tammina2019transfer} or residual networks (ResNet)~\cite{reddy2019transfer} as the foundation to identify a specific object.  The pretrained \stcd, used in our work, was trained with 10K samples for Java-C$\#$ translation~\cite{structcoder_github}.

\BfPara{Tokenization and Embedding} After tokenizing a code input sequence, the code tokens are embedded in $\mathbb{R}^d$, among special tokens. The leaf ($l$) of the code is embedded with the path from the root of the leaf in an abstract syntax tree (AST), and data flow graph (DFG) variables that follow the default embedding will be used for structure-aware self-attention encoding. We note that the tokenization and embedding processes for source and target codes are identical. In \ours, however, we feed the encoder and decoder pipelines with source and target codes with the same programming language and different stylistic features for ``teaching'' the decoder the code transformation task. 

\BfPara{Fine-tuning: Encoder and Decoder} Because \stcd was originally designed to translate Java to C$\#$ (or C$\#$ to Java) and takes the structural information of a code, such as AST and DFG, the encoder/decoder parsers should be modified to extract the AST information for C++ codes correctly. As such, we customized an open-source {\em tree-sitter-cpp} parser~\cite{treesitter}. Moreover, we modified and added the DFG structure to the C++ code format for both the encoder and decoder. To optimize the performance, we fine-tuned the maximum length of the source input, DFG, and AST, and the maximum depth of AST. 
Unlike \stcd, which deals with a small piece of code (\ie functions), \ours's training goal is to work with the whole program at once. This requires not only increasing the maximum input and output length of the T5 model to $1024$ but also increasing the maximum lengths of the AST and DFG to $1000$ to keep pace.

\begin{table}[t]
    \centering
    \begin{tabular}{lccc}
        \Xhline{2\arrayrulewidth}
                                \ccd{}& \ccd{MAA~\cite{quiring2019misleading}}   & \ccd{} & \ccd{\ours{}} \\
        \Xhline{2\arrayrulewidth}
            Number of samples       & 320                &            & Training: 640 \& Testing: 320              \\ 
            \cc{Number of outputs}        & \cc{320}        & \cc{  }                   & \cc{320}                        \\       \hline              
            Processing time                & $\approx$ 11,500 mins      &          & Training: $\approx$ 3,600 mins \& Testing: 58 mins       \\
            \cc{Transformation success rate}        & \cc{97.5\%}           & \cc{}             &  \cc{85\% }                    \\ 
            
        \Xhline{2\arrayrulewidth}
\end{tabular}
    \caption{ A performance comparison: MAA vs. \ours{}. }
    \label{tab:performance}\vspace{-7mm}
\end{table}

\section{Evaluation}\label{sec:evaluation} 
We evaluate the performance of \ours on code obfuscation for misleading code authorship attribution and compare it with MAA~\cite{quiring2019misleading}. In the following, we first explain the experiment setup, dataset, and goals.  Then, we compare the processing time and transformation success rate of \ours with MAA. In addition, we examine the evasion success rates for both Abuhamad \etal~\cite{AbuhamadAMN18} and Caliskan-Islam \etal~\cite{IslamHLNVYG15} methods. Finally, we provide a detailed analysis of syntax and semantic errors that we encountered.

\subsection{Experimental Setup and Goal}
\BfPara{Experiment Setup} We conducted our experiment on a workstation equipped with Nvidia RTX A6000 48GB GPU, Intel Core i7-8700K CPU, and Ubuntu 20.04.5 LTS. 

\BfPara{Dataset} We used the same dataset from Quiring \etal~\cite{quiring2019misleading}, which consisted of C++ files from the 2017 Google Code Jam (GCJ) programming competition~\cite{GoogleCodeJam}. This dataset contains a total of 204 authors and eight challenges (C++ codes) per author (a total of 1,632). The GCJ competition features multiple rounds with multiple participants solving the same programming challenges, allowing us to train a classifier for attribution that concentrates on stylistic patterns rather than artifacts from different challenges. Moreover, for each challenge, there are test sample inputs that can be used to validate the program semantics and sample outputs. 

\BfPara{Untargeted Transformation} The untargeted transformations generate codes predicted as written by any author other than the original author. For evasion, we consider the untargeted transformations due to the structure of \ours{}. This transformation is defined as follows: $f\left(\text{\stcd}(<x_{ij}, x_{i(j+1)}>\right) = y^{*}$ where $y^*$ is anyone other than the original author $y^s$ ($y^* \ne y^s$).



\subsection{Results and Analysis} 
To evaluate the performance of our model against MAA~\cite{quiring2019misleading}, we implemented MAA, Abuhamad \etal~\cite{AbuhamadAMN18}, and Caliskan-Islam \etal~\cite{IslamHLNVYG15}{'s} methods. 
To implement those methods, we utilized the same codes that Quiring \etal~\cite{quiring2019misleading}, available on their GitHub~\cite{code-imitator_github}. 

\begin{table}[t]
\begin{minipage}{.5\linewidth}
    \centering
    \begin{tabular}{lcc}
        \Xhline{2\arrayrulewidth}
            \ccd{Error}           & \ccd{Count} & \ccd{Percentage} \\
        \Xhline{2\arrayrulewidth}
            Undeclared variable     & 10   & 50\%       \\ 
            \cc{Re-declared variable}    & \cc{5}    & \cc{12\%}       \\
            Missing ; or \}         & 4    & 10\%        \\ 
            \cc{Return statement}        & \cc{3}    & \cc{8\%}        \\ 
            Others                  & 8    & 20\%        \\ 
            \hline
            \cc{Total}          & \cc{40}    & \cc{}        \\ 
        \Xhline{2\arrayrulewidth}
\end{tabular}
    \caption{List of syntax errors.}
    \label{tab:syntaxerror}\vspace{-7mm}
\end{minipage}%
\begin{minipage}{.5\linewidth}
    \centering
    \begin{tabular}{lcc}
        \Xhline{2\arrayrulewidth}
            \ccd{Error}           & \ccd{Count} & \ccd{Percentage} \\
        \Xhline{2\arrayrulewidth}
            Misused variable    & 3    & 37\%       \\ 
            \cc{Output statement}    & \cc{4}    & \cc{50}\%       \\
            Input statement     & 1    & 13\%       \\
            \hline
            \cc{Total}          & \cc{8}    & \cc{}        \\ 
        \Xhline{2\arrayrulewidth}
\end{tabular}
    \caption{List of semantic errors.}
    \label{tab:semanticerror}\vspace{-7mm}
\end{minipage}
\end{table}

\BfPara{Processing Time} We measured the processing time for each method, showing that MAA had a drawback, as it took $\approx$11,500 minutes (or 8 days) to generate 320 transformed codes in comparison to \ours{'s} $\approx$ 3,600 minutes (or 2.5 days) for training and 58 minutes for testing, resulting in a 68\% reduction as shown in~\autoref{tab:performance}. In addition to reducing processing time, \ours also has the advantage of not requiring retraining when new features appear. This means that once the model is trained, it can generate 320 transformed codes within only one hour, making it a more efficient and practical solution for misleading code authorship attribution. Moreover, the reduction in processing time not only saves computational resources but also allows for a more efficient and flexible workflow, as it reduces the time required to generate transformed codes and makes it possible to perform misleading code authorship attribution on larger datasets and more complex problems.

\BfPara{Transformation Success Rate} We analyzed the transformation success rate of both methods. In this context, transformation success rate refers to the percentage of transformed codes that are semantically equivalent to the original code and syntactically correct.
Our experiment showed that the MAA had a 97.5\% transformation success rate for 320 samples, with 2.5\% (8) instances of semantic errors. On the other hand, our proposed model performed slightly lower with an 85\% (272) transformation success rate as shown in~\autoref{tab:performance}. We found 12.5\% (40) syntax errors and 2.5\% (8) semantic errors in the outputs from our method. However, it is important to note that while \ours{'s} success rate was lower than MAA{'s}, it still achieved high accuracy while using significantly fewer computational resources.
Moreover, the \ours{'s} syntax errors and semantic errors were minor, with the potential to improve \ours{'s} performance by fine-tuning and using a larger dataset.
In summary, \ours is practical since it achieves comparable results while saving computational resources and maintaining performance stability over different settings.

\BfPara{Syntax Error} Most of the errors we encountered were syntactical errors. However, fortunately, these errors were generally minor and could be easily corrected. The syntax errors we encountered consisted of five types: undeclared variables (10), re-declared variables (5), missing semicolons or curly braces (4), incorrect placement of return statements (3), and other errors (8) as shown in~\autoref{tab:syntaxerror}. 

\BfPara{Semantic Error} We encountered a small portion of semantic error compared to the syntax error. This produces different outputs with the original code under the same input and this error consists of two types of error: misused variable (3), and missing input (or out) statement (1, 4) as shown in~\autoref{tab:semanticerror}. It is important to note that semantic errors are more difficult to detect than syntax errors, as they do not break the code, but they can change the output of the code. In this experiment, we found that semantic errors were much less than syntax errors. However, semantic errors are still important to consider, as they can significantly change the output of the code, even if the code appears to run without any errors.

\BfPara{Evasion Attack Success Rate} The success rate of evasion attacks is calculated as the percentage of instances where the attribution method assigns an incorrect label to a given code. 
To evaluate the effectiveness of \ours{} and MAA's, we start by presenting the baseline accuracies of Abuhamad \etal~\cite{AbuhamadAMN18} and Caliskan-Islam \etal~\cite{IslamHLNVYG15} in Table~\ref{tab:baseline}. However, the list of authors that Quiring \etal used for their work and the authors who were randomly selected by their program for our experiment are different. Therefore, the accuracy that we got from our experiments is slightly different from the results stated in their paper~\cite{quiring2019misleading}. 
\ours{} recorded a success rate of 95.77\% for Abuhamad \etal{'s} and 88.74\% for Caliskan-Islam \etal{'s} methods.
Meanwhile, the success rate of the MAA's was 99.1\% and 99.2\% for Abuhamad \etal{'s} and Caliskan-Islam \etal{'s}, respectively. The evasion attack success rate results are presented in Table~\ref{tab:evasion}. 

\begin{table}[t]
    \centering
    \begin{tabular}{lcc}
        \Xhline{2\arrayrulewidth}
        \ccd{Method}                & \ccd{MAA~\cite{quiring2019misleading}} & \ccd{\ours{}}     \\
        \Xhline{2\arrayrulewidth}
         Abuhamad \etal~\cite{AbuhamadAMN18}             &  99.1\%      &  95.77\%                \\ 
         \cc{Caliskan-Islam \etal~\cite{IslamHLNVYG15}}  & \cc{99.2\%}  &  \cc{88.74\%}      \\            
            
        \Xhline{2\arrayrulewidth}
\end{tabular}
    \caption{Evasion attack success rate of MAA and \ours{} for Abuhamad \etal and Caliskan-Islam \etal methods under untargeted transformation scenario.}
    \label{tab:evasion}\vspace{-7mm}
\end{table}

\vspace{-4mm}
\section{Conclusion}\label{sec:conclusion} 
In this paper, we presented a practical approach for obscuring code authorship attribution by using a machine learning-based Seq2Seq model, called \stcd. We chose this method as the current approach to using MCTS, although it is reliable, has limitations in terms of resource management. Thus, we utilized \stcd to mislead code authorship attribution while reducing resource usage and processing time as well as preserving the performance of the code transformation. As a result, our findings showed that \stcd which was fine-tuned with our dataset significantly reduced processing time while preserving transformation performance.

\end{document}